\documentclass[aps,prl,twocolumn,superscriptaddress]{revtex4}%
\usepackage{amsfonts}
\usepackage{amsmath}
\usepackage{amssymb}
\usepackage{graphicx}
\usepackage{xcolor}
\usepackage{epstopdf}
\usepackage{bm}%
\begin{document}
\title{Phase-manipulation-induced Majorana Mode and Braiding Realization in
Iron-based Superconductor Fe(Te,Se)}
\author{Rui Song}
\affiliation{HEDPS, Center for Applied Physics and Technology and School of Physics, Peking
University, Beijing 100871, China}
\affiliation{HEDPS, Center for Applied Physics and Technology and School of Engineering,
Peking University, Beijing 100871, China}
\affiliation{Anhui Key Laboratory of Condensed Matter Physics at Extreme Conditions, High
Magnetic Field Laboratory, HFIPS, Anhui, Chinese Academy of Sciences, and
University of Science and Technology of China, Hefei, China}
\author{Ping Zhang}
\email{zhang_ping@iapcm.ac.cn}
\affiliation{School of Physics and Physical Engineering, Qufu Normal University, Qufu
273165, China}
\affiliation{HEDPS, Center for Applied Physics and Technology and School of Engineering,
Peking University, Beijing 100871, China}
\affiliation{Institute of Applied Physics and Computational Mathematics, Beijing 100088, China}
\affiliation{Beijing Computational Science Research Center, Beijing 100084, China}
\author{Ning Hao}
\email{haon@hmfl.ac.cn}
\affiliation{Anhui Key Laboratory of Condensed Matter Physics at Extreme Conditions, High
Magnetic Field Laboratory, HFIPS, Anhui, Chinese Academy of Sciences, and
University of Science and Technology of China, Hefei, China}

\begin{abstract}
Recent experiment reported the evidence of dispersing one-dimensional Majorana
mode trapped by the crystalline domain walls in FeSe$_{0.45}$Te$_{0.55}$.
Here, we perform the first-principles calculations to show that iron atoms in
the domain wall spontaneously form the ferromagnetic order in line with
orientation of the wall. The ferromagnetism can impose a $\pi$ phase
difference between the domain-wall-separated surface superconducting regimes
under the appropriate width and magnetization of the wall. Accordingly, the
topological surface superconducting state of FeSe$_{0.45}$Te$_{0.55}$ can give
rise to one-dimensional Majorana modes trapped by the wall. More
interestingly, we further propose a surface junction in the form of
FeSe$_{0.45}$Te$_{0.55}$/ferromagnet/FeSe$_{0.45}$Te$_{0.55}$, which can be
adopted to create and fuse the Majorana zero modes through controlling the
width or magnetization of the interior ferromagnetic barrier. The braiding and
readout of Majorana zero modes can be realized by the designed device. Such
surface junction has the potential application in the superconducting
topological quantum computation.

\end{abstract}
\maketitle

Majorana zero modes (MZMs) have attracted intense attention in condensed
matter physics for the potential application in topological quantum
computation\cite{quancomput-1,quancomput-2,quancomput-3,quancomput-4,quancomput-5,quancomput-6}%
. The paradigm for pursuing them lies in inducing Cooper pairs to the
spin-momentum-locked bands through superconducting proximity effect in the
artificial physical systems. The candidate platforms mainly includes the
topological insulator, the Rashba-spin-orbit-coupled semiconductor and the
ferromagnetic atomic chains forming the heterostructures with the conventional
s-wave superconductors
\cite{proposal-1,proposal-2,proposal-3,proposal-4,proposal-5}. However, the
weakness of the proximity-effect-induced superconductivity requires the
ultra-low temperature and limits the potential applications.

Recently, the electronic bands with nontrivial topology have been
theoretically predicted and experimentally verified in some iron-based
superconductors\cite{bandtopy-1,bandtopy-2,bandtopy-3,bandtopy-4,bandtopy-5,bandtopy-6}%
. The coexistence of topology and high-temperature superconductivity make the
iron-based superconductors as the unique platform to realize the
high-temperature topological
superconductors\cite{bandtopy-4,sctopy-1,sctopy-2,sctopy-3,sctopy-4,hightopy-1,hightopy-2,hightopy-3,hightopy-4,hightopy-5}%
. In general, some defects can trap the MZMs in topological\ superconductors.
Assisted by angle resolved photoelectron spectroscope (ARPES), the scanning
tunnel microscope/spectrum (STM/S) has observed the zero-energy conductance
anomalies located on some point defect such as iron impurity and
superconducting vortex in Fe(Se,Te), Li$_{1-x}$Fe$_{x}$HOFeSe, CaKFe$_{4}%
$As$_{4}$, and Fe(Se,Te)/STO, which strongly indicate the emergence of
MZMs\cite{pointmajor-1,pointmajor-2,pointmajor-3,pointmajor-4,pointmajor-5,pointmajor-6,line-df-1,line-df-2,line-df-3}%
. Unlike the point-like MZMs, more recently, an unexpected one-dimensional
(1D) dispersive Majorana mode trapped by the crystalline domain walls (DWs) in
FeSe$_{0.45}$Te$_{0.55}$ have been experimentally reported\cite{major-1d}. It
is found that the DWs and 1D Majorana mode show some interesting features. For
instance, the lattices form the bulge structure approaching the DWs. The
orientation of the DW has a deflection of about a 45 degree angle against the
direction of lattice shift. The differential conductance $dI/dV$ spectra show
some subtle differences at the different positions of DWs. However, the
physical origin and the features of the 1D Majorana mode have not been
comprehensively understood.

In this work, we perform the first-principles calculations to investigate the
properties of the DWs in FeSe$_{0.45}$Te$_{0.55}$. The numerical results can
capture both the bulge structure and the specific orientation of the DWs.
Interestingly, we find that the iron atoms in the DWs spontaneously form the
ferromagnetic order with the magnetization direction along the DWs.
Accordingly, we show that the ferromagnetism can manipulate the surface
superconducting phase difference between the two sides of the DWs, which is
analogous to the physical picture of Fulde-Ferrell-Larkin-Ovchinnikov (FFLO)
superconducting state\cite{FFLO-1,FFLO-2}. The appropriate width and
magnetization of the DWs can give $\pi$ phase difference and induce 1D
Majorana mode. The fluctuations of the width and magnetization of the DWs make
the phase difference to slightly deviate from $\pi$, which results in the
subtle differences between the position-dependent $dI/dV$ spectra. More
meaningfully, we further propose a surface junction in the form of
FeSe$_{0.45}$Te$_{0.55}$/ferromagnet/FeSe$_{0.45}$Te$_{0.55}$ (S/F/S). Tuning
the width or magnetization in the different regimes of the ferromagnet, the
surface junction can be adopted to create and fuse the MZMs located on the
boundary between different regimes. The braiding and readout of MZMs can be
realized by the designed device based on such junction.

\begin{figure}[pt]
\begin{center}
\includegraphics[width=1.0\columnwidth]{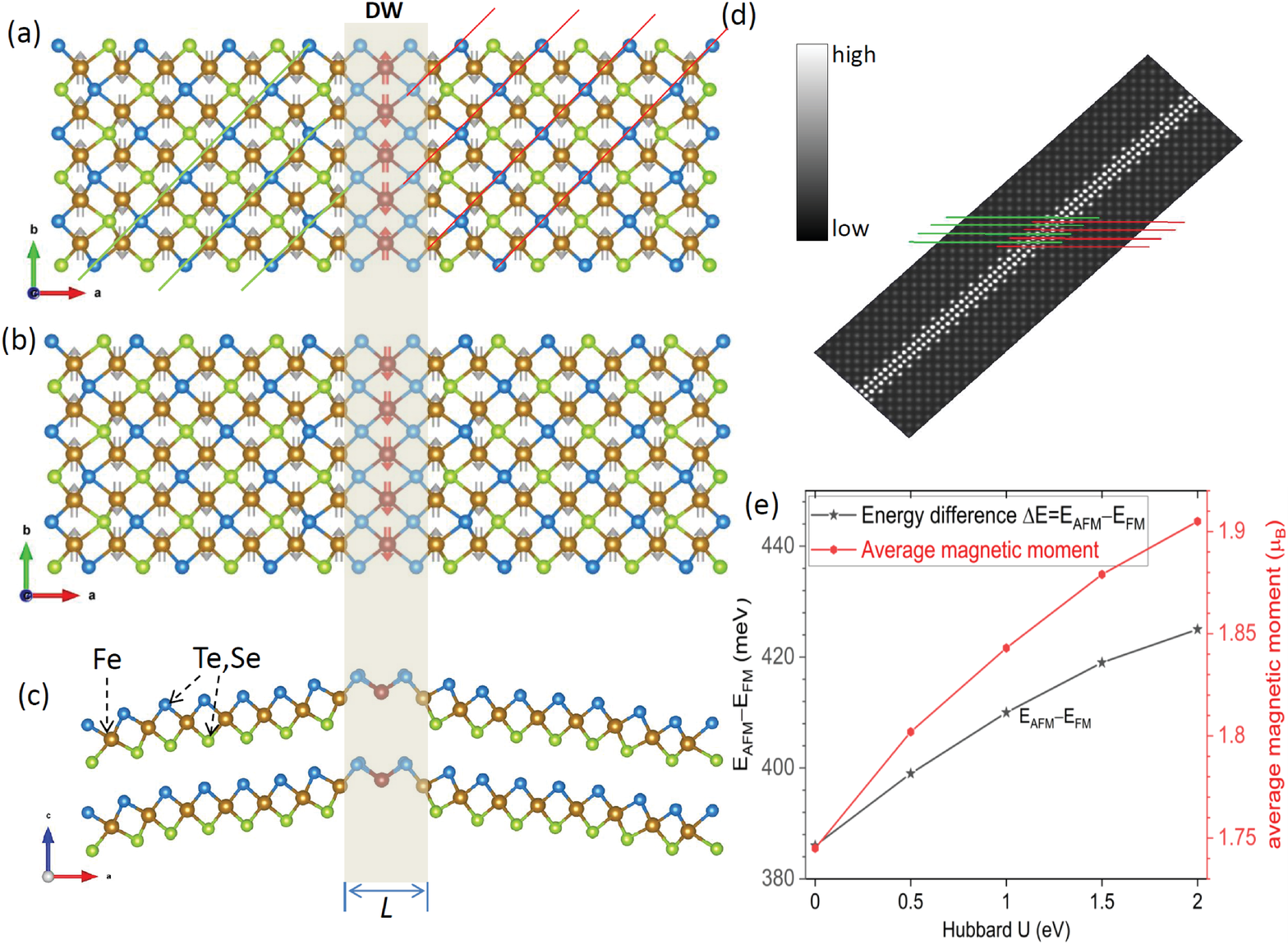}
\end{center}
\caption{(a)-(c) Top and side view of Fe(Te,Se) with 14$\times$1$\times$1
super cell with DW labeled by the shadowed region. In both (a) and (b), the
two sides of the DW have the collinear AFM order, but the iron atoms have the
AFM and FM order in the DW in (a) and (b), respectively. (d) The simulated STM
pattern. (e) The magnetic moment and the energy difference between the
magnetic patterns in (a) and (b), i.e., $E_{AFM}-E_{FM}$.}%
\label{figlda}%
\end{figure}

As the STM experiment shows, the DWs are a line of dislocation-like defects,
which separate the lattices into left and right parts with a relative
half-unit-cell shift\cite{major-1d}. The thickness of DWs is one or several
unit cells of Fe(Te,Se) from experimental observation\cite{DW-thickness}.
Therefore, we construct a $14\times1\times1$ supercell of Fe(Te,Se), and the
left and right seven unit cells are connected by a horizontal mirror
reflection about iron plane. The calculation details are present in Ref.
\cite{SM} . The stable structure is displayed in Fig. \ref{figlda} (a)-(c),
with a obvious bulge structure deformation. The simulated STM image in Fig.
\ref{figlda} (d) reveals that the DWs are much lighter than other regions in
accord with the experimental observations\cite{SM}. Furthermore, the
deflection of about a 45 degree angle of the orientation of the DW against the
direction of half-unit-cell shift is also captured by the simulation in
comparison with image in Fig. 3(d) of Ref.\cite{major-1d}. The bulge structure
deformation is due to the unsymmetric Fe-Te(Se) chemical bonds in the DWs,
which break the force balance along $\mathbf{z}$ direction. Another impact
from the unsymmetric bonds is to enhance the localization of irons atoms in
the DWs, which have a tendency to generate the magnetic moments. To determine
the possible magnetic order of the DWs, for convenience, we still assume the
Fe(Te,Se) at the two sides of DWs possess the antiferromagnetic (AFM) order in
spite of only strong AFM fluctuation existing in Fe(Se,Te). This is the
general strategy adopted by the calculations to determine the magnetism of
iron-based superconductors\cite{magnetism-1,magnetism-2,magnetism-3}. It is
well accepted that the superconducting state in Fe(Te,Se) has a close
relationship with the collinear AFM order, which we consider as shown in Fig.
\ref{figlda} (a) and (b). For the iron atoms in the DWs, we consider the AFM
and FM pattern in Fig. \ref{figlda} (a) and (b), respectively. The
ground-state magnetic pattern is self-consistently calculated and determined.
As shown in Fig. \ref{figlda} (e), the magnetic pattern in Fig. \ref{figlda}
(b) has lower energy than that in Fig. \ref{figlda} (a). The Hubbard U can
further enlarge the energy difference. The calculated results are not
unexpected, because the position switchings of the anions do not dramatically
modulate the amplitudes of the nearest and next nearest neighbor exchange
couplings. According to the $J_{1}$-$J_{2}$
model\cite{magnetism-3,Exchange-1,Exchange-2}, the magnetic pattern shown in
Fig. \ref{figlda} (b) is reasonable. Note that we also calculate the lattice
pattern with domain walls involving two lines of irons. The same ferromagnetic
order is also obtained\cite{SM}. Note that the local ferromagnetism in
Fe(Te,Se) was verified by recent experiments\cite{FM-evidence1,FM-evidence2}.

\begin{figure}[pt]
\begin{center}
\includegraphics[width=1.0\columnwidth]{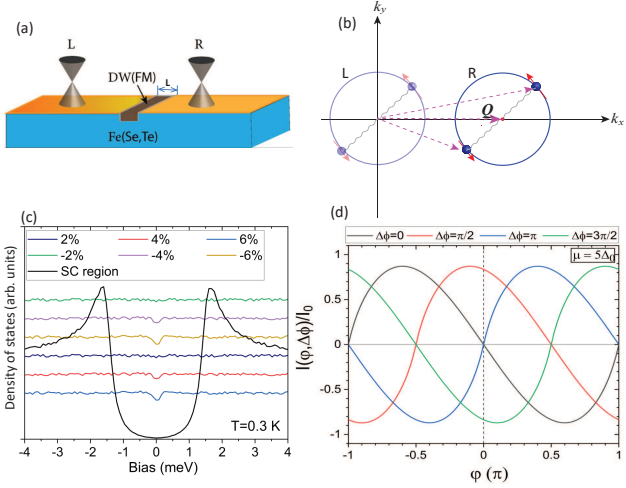}
\end{center}
\caption{(a) Schematic plotting for the surface S/F/S junction. The left (L)
and right (R) gray Dirac cones denote the topological surface bands of
Fe(Te,Se), and the middle labels the ferromagnetic DW. (b) The Fermi surface
of the R Dirac cone shifts momentum $Q$ in the reference of L Dirac cone. The
pairing is indicated by the wave lines. (c) The calculated density of \ states
(DOS) for the different $\Delta\phi$ fluctuations [-6\%, 6\%] about $\pi$. The
U-shape DOS is for the uniform superconducting state. The temperature is 0.3K.
(d) The oscillation of tunneling current $I(\varphi,\Delta\phi)$ about
$\varphi$ for different $\Delta\phi$. Note that the current is not zero for
$\varphi=0$ and $\Delta\phi\neq0,\pi$, which is different from the
conventional junction. }%
\label{figsfs}%
\end{figure}

Below superconducting transition temperature $T_{c}$, the superconductivity
arises through suppressing the collinear AFM order or spin fluctuation in
normal state, while the FM order in the DWs is preserved due to the absence of
superconductivity. Then, the system can be simplified as a S/F/S junction
which only involving the topological surface states, as shown in Fig.
\ref{figsfs}(a). The ferromagnetism of the DWs could generate an exchange
field, \textit{i.e.}, $h_{y}=(n\langle S_{i}^{y}\rangle/\mu_{B})\int
J(\mathbf{r})d^{3}r$, with $n$, $S_{i}^{y}$, $\mu_{B}$, and $J(\mathbf{r})$
labeling the concentration of localized moments, the average value of the
localized spins, Bohr magneton and ferromagnetic exchange integral,
respectively. Hereafter, we set $\mu_{B}=1$ for convenience. It has been shown
that the F layer can generally modulate a superconducting phase difference
between the two S layers. Here, we give a simple picture to address this
point. Consider the normal Fermi surface of both S layer described by the
topological surface Hamiltonian $H_{surf}=v_{F}(\mathbf{\sigma}\times
\mathbf{k})\cdot\hat{z}$, the Fermi surface of the right S layer is shifted
$\mathbf{Q}=h_{y}/v_{F}\mathbf{\hat{k}}_{x}$ by the effective Zeeman term
$g\sigma_{y}h_{y}/2$ in reference to the one of the left S layer. From the
first-principles calculations, we estimate the effective Zeeman energy
$h_{y}\sim55$ meV\cite{SM}. The Fermi velocity $v_{F}\sim216\ $meV\AA \ and
$k_{F}\sim0.03$\AA $^{-1}$\cite{Fermienergy}. Thus $Q\sim0.25$\AA $^{-1}$, and
$Q\gg$ $k_{F}$. The left and right Fermi surfaces are fully separated, as
shown in Fig. \ref{figsfs}(b). The superconducting order parameter
$\Delta(\mathbf{Q})$ is proportional to $\langle\hat{c}_{\mathbf{k}\uparrow
}\hat{c}_{2\mathbf{Q}-\mathbf{k}\downarrow}\rangle$, which becomes spatially
oscillating after Fourier transform into the real space. Namely,
$\Delta(\mathbf{R})=\Delta_{0}e^{i2\mathbf{Q}\cdot\mathbf{R}}$ with
$\Delta_{0}$, $\mathbf{R}$ labeling the induced superconducting order
parameter of the topological surface band and the coordinate of center of mass
of Cooper pair, respectively. This is equivalent to stating that a Cooper pair
propagates across a Zeeman field and acquire a finite momentum and a phase
oscillation in the real space, which is similar to the FFLO state
\cite{FFLO-1,FFLO-2}. The phase shift is related to the width $L$ and
intensity of the effective Zeeman field $h_{y}$ of DWs and can be expressed as
$\Delta\phi=2h_{y}L/v_{F}$. When the system is in equilibrium state, no random
phase difference occurs between the left and right superconducting regions,
however, the system becomes a surface $\Delta\phi$-junction due to the
phase-shift effect. According to Fu and Kane's theorem \cite{proposal-1}, a 1D
Majorana mode will arise in the DWs when $\Delta\phi=\pi$ in the limit of
$L\rightarrow0$. 

To verify the argument, we consider the generic surface S/F/S junction with
arbitrary width $L$ and effective Zeeman field $h_{y}$ of DWs, and solve the
Dirac-Bogoliubov-de Gennes (BdG) equation,
\begin{equation}
H_{BdG}\Psi=E\Psi\ , \label{D-BdG}%
\end{equation}
with
\begin{equation}
H_{BdG}=\left(
\begin{array}
[c]{cc}%
H_{0}(k) & i\sigma_{y}\Delta(r)\\
-i\sigma_{y}\Delta^{\ast}(r) & -H_{0}^{\ast}(-k)
\end{array}
\right)  . \label{D-BdG-1}%
\end{equation}
Here, $H_{0}(k)=v_{F}(\sigma_{x}k_{y}-\sigma_{y}k_{x})-\mu+\sigma_{y}%
h_{y}\Theta(\frac{L}{2}-|x|)$ and $\Delta(r)=\Delta_{0}e^{-i\frac{\varphi}{2}%
}\Theta(-x-\frac{L}{2})+\Delta_{0}e^{i\frac{\varphi}{2}}\Theta(x-\frac{L}{2}%
)$, and $\Theta$ is the Heaviside step function, $\varphi$ is the random phase
difference between two surface S regimes. The details of the analytic
calculation of Eq. (\ref{D-BdG}) are shown in Ref. \cite{SM}. Here, we only
discuss the main results. For both $\mu\rightarrow0$ and $\mu\gg\Delta_{0}$,
two branches of bound states have the same dispersion,
\begin{equation}
\varepsilon(k_{y})=\pm\sqrt{v_{F}^{2}k_{y}^{2}+\Delta_{0}^{2}\cos^{2}%
(\frac{\varphi}{2}-\frac{h_{y}L}{v_{F}})}. \label{dispersion}%
\end{equation}
Here, $k_{y}\rightarrow0$ for $\mu\rightarrow0$, and $k_{y}=\mu/v_{F}%
\sin\theta$ with $\theta$ the incident angle approaching 0. In equilibrium
state with $\varphi=0$, a couple of gapless bound states can be obtained when
$\frac{h_{y}L}{v_{F}}=\frac{\Delta\phi}{2}=\frac{(2n+1)\pi}{2}$ with $n$ the
integer number. Namely, 1D dispersive Majorana modes with constant density of
states (DOS) are realized for the FM DWs with appropriate width $L$ and
magnetization $h_{y}$. For finite $\mu$, the effective low-energy Hamiltonian
$H_{eff}$ describing a couple of bound states can be obtained by projecting
$H_{BdG}$ into the subspace spanned by two eigen-states $\xi_{\pm}$
corresponding to $\varepsilon(k_{y})=0$ with $k_{y}=0$, $\varphi=0$,
$\frac{h_{y}L}{v_{F}}=$ $\frac{\pi}{2}$\cite{proposal-1}, and $\xi_{\pm}%
=\frac{1}{2}(\mp i,-1,\pm i,1)^{T}e^{\pm i\mu x/v_{F}-\int_{0}^{|x|}d\tilde
{x}\Delta_{0}(\tilde{x})/v_{F}}$. Then,%

\begin{equation}
H_{eff}=\tilde{v}_{F}q_{y}\tau_{y}-\Delta_{0}\cos(\Delta\phi/2)\tau_{z}.
\label{D-BdG-2}%
\end{equation}
Here, $\tilde{v}_{F}=v_{F}[\cos k_{F}L+(\Delta_{0}/\mu)\sin k_{F}L]\Delta
_{0}^{2}/(\mu^{2}+\Delta_{0}^{2})$ with $\tau_{y}$ and $\tau_{z}$ the Pauli
matrices spanned by particle-hole space. When $\Delta\phi=\pi$, $H_{eff}$ in
Eq. (\ref{D-BdG-2}) still gives the 1D dispersive Majorana modes with constant DOS.

The implication for the experimental observations\cite{major-1d} can be
elucidated as follows. The calculated effective Zeeman energy $h_{y}\sim55$
meV. The width of DW is about twice about the lattice constant $L\sim
6.2\ $\AA . The Fermi velocity $v_{F}\sim216\ $meV\AA \cite{Fermienergy}. Thus
the phase shift can be derived as $\Delta\phi\sim\pi$, which indicates that
the 1D dispersive Majorana modes in Fe(Te,Se) is likely to be induced due to
the ferromagnetism of DWs. However, the realistic sample could not meet such
rigorous condition of $\Delta\phi=\pi$. There should exist some slight
fluctuations of $h_{y}$ and $L$ in the sample. Accordingly, $\Delta\phi$
should have tiny fluctuations around $\pi$. As a result, the experiment indeed
observed the $dI/dV$ spectra showed some subtle differences at the different
positions of DWs\cite{major-1d}. Such behavior is captured by the
calculation\cite{SM}, as shown in Fig. \ref{figsfs} (c). Besides, the picture
of such surface S/F/S junction proposed here can also be verified by measure
the tunneling current, which can be expressed as,%

\begin{equation}
I(\varphi,\Delta\phi)=-\frac{eN}{\hbar}\int_{-\pi/2}^{\pi/2}\frac
{\partial|\varepsilon(k_{y})|}{\partial\varphi}\tanh\frac{|\varepsilon
(k_{y})|}{2k_{B}T}\cos\theta d\theta. \label{current}%
\end{equation}
Here, $k_{B}$ is Boltzmann constant. $N=k_{F}W/\pi$ denoting the number of the
transport channels in the system with $W$ the length of the domain wall. As
shown in Fig. \ref{figsfs} (d), the tunneling current $I(\varphi,\Delta\phi)$
is none zero for $\varphi=0$ and $\Delta\phi\neq0$ and $\pi$. Thus, the
surface states of FeSe$_{0.45}$Te$_{0.55}$ with domain wall form a $\Delta
\phi$ phase battery\cite{phase-batt,phase-batt-1}.

\begin{figure}[pt]
\begin{center}
\includegraphics[width=1.0\columnwidth]{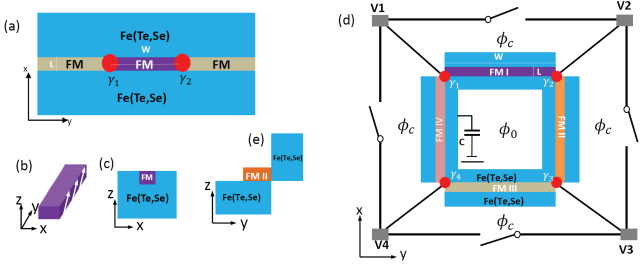}
\end{center}
\caption{(a) A basic device to create and fuse a pair of MZMs through tuning
the magnetization of different regimes of the FM film. (b) The magnetic
moments (white arrows) of the FM film can be tuned in the y-z plane by some
external field. (c) The cross section of the middle part in (a). (d) The
extended device built from the basic device. Four FM films I, II, III, IV are
labeled. $\phi_{0}$ denotes the phase difference between the inner
superconductor loop and the four outer superconductors. The superconductor is
connected to ground by a capacitor. (e) The cross section of left edge of the
square in (d).}%
\label{figbrd}%
\end{figure}

According to Eq. (\ref{D-BdG-2}), the boundary of the inverse masses can trap
a MZM\cite{proposal-1}. Thus, the surface S/F/S junction can be upgraded by
replacing the DWs with a thin FM film, as shown in Fig. \ref{figbrd} (a).
According to $\Delta\phi=2h_{y}L/v_{F}$, there are two ways to realize MZMs.
The first one is to tune $h_{y}$ of different regimes of F. In such case, the
FM film can be fabricated with the soft magnetic materials, whose
magnetization can be easily tuned by the electric, magnetic field etc., as
shown in Fig. \ref{figbrd} (b). Note that the $z$-direction magnetization
energy smaller than chemical potential and superconducting gap does not change
the results. The second one is to tune $L$ of different regimes of F, where
the hard magnetic materials can be adopted to fabricate the FM film due to the
robust magnetization. In both cases, the candidate magnetic materials could be
selected among the yttrium iron garnet (YIG) ferrite, which includes many soft
and hard ferrites such as (MnZn)O$\cdot$Fe$_{2}$O$_{3}$ and BaO$\cdot6$%
Fe$_{2}$O$_{3}$, and the relevant film technique is very mature. Here, we
adopt the first way, and the basic device is shown in Fig. \ref{figbrd} (a)
and (c). Then, the mass term $m(h_{y})=\Delta_{0}\cos(h_{y}L/v_{F})$. Through
tuning $h_{y}$ to have $m(h_{y})<0$ in the middle regime of F and $m(h_{y})>0$
in two side regimes, a pair of MZMs can be created at the boundaries between
the middle and side regimes. They can also fuse by adiabatically tuning
$m(h_{y})$ to have same sign in all regimes. Namely, such basic device can
realize the creation and fusion of a pair of MZMs through selectively tuning
the magnetization $h_{y}$ of F. \begin{figure}[ptpt]
\begin{center}
\includegraphics[width=1.0\columnwidth]{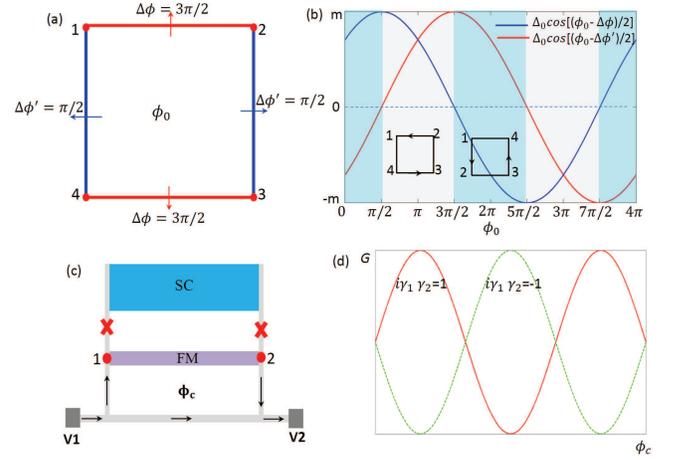}
\end{center}
\caption{ (a) The square denotes the loop formed by the four FM films.
$\Delta\phi=3\pi/2$ and $\Delta\phi^{\prime}=\pi/2$ denote the values of
$2h_{y}L/v_{F}$ in regime I, III and II, IV, respectively. (b) The evolution
of two masses $m^{(\prime)}=\Delta_{0}\cos[(\phi_{0}-\Delta\phi^{(\prime
)})/2]$ as a phase difference $\phi_{0}$. The insert labels the equivalent
exchange process of MZMs during the braiding. (c) The equivalent interference
circuit to readout fermion parity of $i\gamma_{1}\gamma_{2}$. Note that the
superconductor path is disconnected by the capacitor. (d) The
interference oscillation of conductance as change as $\phi_{c}$ for different
fermion parity. }%
\label{figbrd1}%
\end{figure}

To realize and detect the braiding of the
MZMs\cite{Fu-detect,Fu-detect1,braiding-1,braiding-2,braiding-3,braiding-4,braiding-5,braiding-6,braiding-7}%
, the basic device has to involving two pairs of MZMs at least. The new device
is shown in Fig. \ref{figbrd} (d) with the configuration of the cross section
in Fig. \ref{figbrd} (e). Then, the superconducting square in Fig.
\ref{figbrd} (e) disconnects with the outer superconductors and the phase
difference is denoted by $\phi_{0}$. First, tuning $h_{y}$ to have
$m(h_{y})<0$ in I, III regimes and $m(h_{y})>0$ in II, IV regimes as shown by
the configuration in Fig. \ref{figbrd1} (a), two pairs of MZMs labeled by
$\gamma_{1}$, $\gamma_{2}$, $\gamma_{3}$, $\gamma_{4}$ are created and located
on the four corners of the square. Two fermions can be defined as
$f_{1/2}=(\gamma_{1/3}+i\gamma_{2/4})/2$. Suppose the initial state is
occupied state of the two fermions, \textit{i.e.},
$\vert$%
11$\rangle=f_{1}^{\dag}f_{2}^{\dag}$%
$\vert$%
00$\rangle$ with
$\vert$%
00$\rangle$ the vacuum state of both fermions. Second, the braiding operation
is realized by adiabatically advancing the phase difference $\phi_{0}$ from 0
to 2$\pi$. During the process, the mass term $m=\Delta_{0}\cos[(\phi
_{0}-\Delta\phi)/2]$ also adiabatically evolves, as shown in Fig.
\ref{figbrd1} (b). This braiding operation is equivalent to exchange
$\gamma_{2}$ and $\gamma_{4}$, \textit{i.e.}, $\gamma_{2}\rightarrow$
$\gamma_{4}$, $\gamma_{4}\rightarrow$ $-\gamma_{2}$, as shown in the insert of
Fig. \ref{figbrd1} (b). Here, we assume the cut line is the right edge of the
square\cite{proposal-1}. The relevant braiding operator is $\mathcal{O}%
_{42}=(1+\gamma_{4}\gamma_{2})/\sqrt{2}$, under which, the final state
$\vert$%
$\Psi_{f}\rangle_{1}=\mathcal{O}_{42}|11\rangle=(|11\rangle-|00\rangle
)/\sqrt{2}$. The twice braidings give
$\vert$%
$\Psi_{f}\rangle_{2}=$ $\mathcal{O}_{42}^{2}|11\rangle=-|00\rangle$, which
means the fermion occupied number changes two. Note that the braiding result
is unchanged by exchanging $\gamma_{1}$ and $\gamma_{3}$ and pinning
$\gamma_{2}$ and $\gamma_{4}$. Third, the braiding results can be readout by
electron teleportation effect of MZMs\cite{Fu-detect,Fu-detect1} with the
interference circuit in Fig. \ref{figbrd} (d). Take a pair of $\gamma_{1}$,
$\gamma_{2}$ as an example. To detect the braiding results of fermion parity
of $i\gamma_{1}\gamma_{2}$, one can connect the switch between $\gamma_{1}$
and $\gamma_{2}$ in Fig. \ref{figbrd} (d) and apply a magnetic flux $\phi_{c}%
$. The equivalent circuit is shown in Fig. \ref{figbrd1} (c). The current
flows along two different paths indicated by the arrows in Fig. \ref{figbrd1}
(c). Note that the superconductor path can be forbidden by tuning the
capacitor in Fig. \ref{figbrd} (d). Accordingly, the measured conductance
should form interference oscillation as change as $\phi_{c}$, \textit{i.e.},
$G=g_{0}+i\gamma_{1}\gamma_{2}g_{1}\cos[e(\phi_{c}-\phi_{i})/\hbar]$ with
$\phi_{i}$ the intrinsic phase difference\cite{Fu-detect1}, and the fermion
parity of $i\gamma_{1}\gamma_{2}$ can be readout, as shown in Fig.
\ref{figbrd1} (d)\cite{Fu-detect,Fu-detect1}. The twice braiding can give the
result of definitive sign change of $i\gamma_{1}\gamma_{2}$ between the
initial and final states.

At last, we discuss the experimental feasibility and advantage of the designed
device. First, the decay length of MZM is $\xi_{0}\sim\tilde{v}_{F}/\Delta
_{0}\sim$ 7\AA \ for $\mu=5\Delta_{0}$, $v_{F}\sim$ 216 meV\AA \ and
$\Delta_{0}\sim$ 1.8 meV\cite{sctopy-4,major-1d,Fermienergy}. The size of MZM
is very local. Second, the other quasi-particle energy is roughly estimated by
$\sim\tilde{v}_{F}\pi/W$\cite{proposal-1}. If the temperature $k_{B}T\sim
0.1$meV. The length of $W$ can be $\sim$ 27nm for $\mu=5\Delta_{0}$ and much
larger than the decay length $\xi_{0}$. When temperature is below $1$ K$\sim$
0.1 meV, the MZMs and relevant braiding are robust against the thermal
fluctuation. The charging energy $U$ $\sim Q_{0}^{2}/C$ is required to
comparable to $\Delta_{0}\sim$ 1.8meV\cite{Fu-detect,Fu-detect1}, which can
lower the difficulty in control of the capacitor.

In conclusion, we propose a surface S/F/S junction to reveal the underlying
physics of experimentally observed dispersive 1D Majorana mode in
FeSe$_{0.45}$Te$_{0.55}$. Our calculations predict the spontaneously
ferromagnetic order in the domain wall, and we prove the crucial role of
ferromagnetism to modulate the superconducting phase of the surface electrons
to form Majorana modes. More meaningfully, we design a feasible artificial
device involving FeSe$_{0.45}$Te$_{0.55}$/ferromagnet/FeSe$_{0.45}$Te$_{0.55}$
junctions, which can create, fuse, braid and readout the Majorana zero modes.

\begin{acknowledgments}
We thank J. P. Hu, X. X. Wu, S. B. Zhang, S. S. Qin, F. W. Zheng, H. F. Du, L.
Shan, Z. Y. Wang, S. C. Yan and X. Y. Hou for helpful discussions. This work
was financially supported by the National Key R\&D Program of China No.
2017YFA0303201, National Natural Science Foundation of China under Grants (No.
12022413, No. 11674331, No.11625415), the \textquotedblleft Strategic Priority
Research Program (B)\textquotedblright\ of the Chinese Academy of Sciences,
Grant No. XDB33030100, the `100 Talents Project' of the Chinese Academy of
Sciences, the Collaborative Innovation Program of Hefei Science Center, CAS
(Grants No. 2020HSC-CIP002), the CASHIPS Director's Fund (BJPY2019B03), the
Science Challenge Project under Grant No. TZ2016001, the Major Basic Program
of Natural Science Foundation of Shandong Province (Grant No. ZR2021ZD01). A
portion of this work was supported by the High Magnetic Field Laboratory of
Anhui Province, China.
\end{acknowledgments}

\end{document}